\documentclass[epsf]{aa}
\usepackage{natbib}
\usepackage{epsfig}

\newcommand{\teff}{\mbox{$T_{\rm eff}$}}
\newcommand{\logg}{\mbox{$\log g$}}

\newcommand{\mictrb}{\mbox{$\xi_{\rm t}$}}
\newcommand{\mactrb}{\mbox{$v_{\rm mac}$}}

\newcommand{\kms}{\mbox{km\,s$^{-1}$}}
\newcommand{\halpha}{\mbox{$H_\alpha$}}
\def\ms{\,m\,s$^{-1}$}         
\def\kms{\,km\,s$^{-1}$}       
\def\teff{$T_{\rm eff}$}
\def\logg{$\log g$}

\def\mictrb{$\xi_{\rm t}$}

\def\kms{km\, s$^{-1}$}

\begin{document}
\title{Discovery and characterization of WASP-6b, an inflated sub-Jupiter mass planet transiting a solar-type star\thanks{Based on data collected with the HARPS spectrograph at ESO La Silla Observatory in the programs 082.C-0040(E) and 082.C-0608(E).}}
\subtitle{}
\author{M. Gillon$^{1,2}$,  D. R. Anderson$^3$,  A. H. M. J.  Triaud$^1$, C. Hellier$^3$,  P. F. L.  Maxted$^3$, D. Pollaco$^4$,  D. Queloz$^1$, B. Smalley$^3$,  R. G. West$^{5}$, D. M. Wilson$^3$, S. J. Bentley$^3$,  A. Collier Cameron$^6$,  B. Enoch$^{6}$, L. Hebb$^6$, K. Horne$^6$, J. Irwin$^7$, 
 Y. C. Joshi$^4$, T. A. Lister$^{8}$,  M. Mayor$^1$, F. Pepe$^{1}$,  N. Parley$^{5}$, D. Segransan$^{1}$,  S. Udry$^1$, P. J. Wheatley$^9$} 
 \offprints{michael.gillon@obs.unige.ch}

\institute{$^1$  Observatoire de Gen\`eve, Universit\'e de Gen\`eve, 51 Chemin des Maillettes, 1290 Sauverny, Switzerland\\
$^2$ Institut d'Astrophysique et de G\'eophysique,  Universit\'e
  de Li\`ege,  All\'ee du 6 Ao\^ut, 17,  Bat.  B5C, Li\`ege 1, Belgium \\
$^3$ Astrophysics Group, Keele University, Staffordshire, ST5 5BG, UK\\
$^4$ Astrophysics Research Centre, School of Mathematics \& Physics, Queen's University, University Road, Belfast, BT7 1NN, UK\\
$^5$ Department of Physics and Astronomy, University of Leicester, Leicester, LE1 7RH, UK\\
$^6$ School of Physics and Astronomy, University of St. Andrews, North Haugh, Fife, KY16 9SS, UK\\
$^7$ Department of Astronomy, Harvard University, 60 Garden Street, MS 10, Cambridge, Massachusetts 02138, USA\\
$^8$ Las Cumbres Observatory, 6740 Cortona Dr. Suite 102, Santa Barbara, CA 93117, USA\\ 
$^9$ Department of Physics, University of Warwick, Coventry, CV4 7AL, UK\\
}

\date{Received date / accepted date}
\authorrunning{M. Gillon et al.}
\titlerunning{The transiting sub-Jupiter mass planet WASP-6b}
\abstract{We report the discovery of WASP-6b,  an inflated sub-Jupiter mass planet transiting every  $ 3.3610060^{+ 0.0000022 }_{- 0.0000035 } $ days a mildly metal-poor solar-type star of magnitude V=11.9.  A combined analysis of the WASP photometry, high-precision followup transit photometry and radial velocities yield a  planetary mass $M_p = 0.503^{+0.019}_{-0.038}$ $M_J$ and radius $R_p = 1.224^{+0.051}_{-0.052}$ $R_J$, resulting in a density  $\rho_p = 0.27 \pm 0.05$ $\rho_J$. The mass and radius for the host star are $M_\ast = 0.88^{+0.05}_{-0.08}$ $M_\odot$ and $R_\ast = 0.870^{+0.025}_{-0.036}$  $R_\odot$.  The non-zero orbital eccentricity $e = 0.054^{+0.018}_{-0.015}$ that we measure suggests that the planet underwent a massive tidal heating $\sim$ 1 Gyr ago that could have contributed to its inflated radius. High-precision radial velocities obtained during a transit allow us to measure a sky-projected angle between the stellar spin and orbital axis $\beta = 11^{+14}_{-18}$ deg. In addition to similar published measurements, this result favors a dominant migration mechanism based on tidal interactions with a protoplanetary disk.
\keywords{binaries: eclipsing -- stars: individual: WASP-6 -- planetary systems -- techniques: photometric  -- techniques: radial velocities -- techniques: spectroscopic} }

\maketitle

\section{Introduction}

Transiting planets play an important role in our understanding of the nature of the extrasolar planetary objects. They are the only exoplanets for which an accurate measurement of the mass and radius is available. The deduced density is a key parameter to constraint theoretical models for the formation, evolution and structure of planets (e.g. Fortney et al. 2007; Liu et al. 2008). For the brightest transiting systems, a study of the atmospheric composition and physics is possible, even with existing instruments like  $HST$ or $Spitzer$ (e.g. Charbonneau et al. 2008; Swain et al. 2008). The discovery rate of transiting planets has increased recently thanks mainly to the efficiency of the CoRoT space-based survey (Baglin et al. 2006) and of a  handful of ground-based wide-field surveys targeting rather bright stars (V $<$ 13):  HATNet (Bakos et al. 2004), WASP (Pollaco et al. 2006), TrES (O'Donovan et al. 2006), and XO (McCullough et al. 2005).

The $\sim$ 50 transiting planets known at the time of this writing show a broad range of mass and radius. Their masses go from 23 $M_\oplus$ for the hot Neptune GJ\,436b (Butler et al. 2004; Gillon et al. 2007) to more than 10 $M_J$ for XO-3 (Johns-Krull et al. 2008). Many planets have a size in concord with basic models of irradiated planets (e.g. Burrows et al. 2007, Fortney et al. 2007), some of them like HD\,149026\,b  (Sato et al. 2005) appearing to be very rich in heavy elements. Nevertheless, a few planets like HD\,209458\,b (Charbonneau et al. 2000, Henry et al. 2000) are `anomalously' large. Several hypothesis have been proposed to explain this radius anomaly, most importantly tides (Bodenheimer et al. 2001; Jackson et al. 2008b),  tides with atmospheric circulation (Guillot \& Showman 2002) and enhanced  opacities (Guillot et al. 2006, Burrows et al. 2007). The existence of several correlations between  parameters of transiting systems has been proposed, for instance between the planet mass and the orbital period (Mazeh et al. 2005;  Gaudi et al. 2005) and between the heavy-element content of the planet and the stellar metallicity (Guillot et al. 2006; Burrows et al. 2007).  The astrophysics  supporting these correlations has still to be fully understood.

It is highly desirable to detect and characterize thoroughly many more bright short period transiting systems to improve our understanding of the highly irradiated gaseous planets and to constraint the structure and evolution models for these objects. With its very high detection efficiency, the WASP transit survey is making a large contribution to this goal. It is the only transit survey operating in both hemispheres: it uses an instrument named WASP-North and located at La Palma to search for planets from the Northern hemisphere and a twin instrument named WASP-South and located at  Sutherland to do the same from the Southern hemisphere. Each of these instruments covers a huge field of view of 482 square degrees per pointing, allowing them to search for transiting planets in a large portion of the sky. Due to the brightness of the host stars, planets detected by WASP are very good targets for high-precision followup observations. For instance, it is possible to measure for most of them the alignment between the stellar rotation axis and the planetary orbital axis via the observation of the Rossiter-McLaughlin effect (RM; Queloz et al. 2000). The measured value for this spin-orbit angle is a strong constraint for inward planetary migration models (see Winn 2008 and references therein). 

We report here the discovery and characterization of WASP-6b, a new sub-Jupiter mass  planet transiting a mildly metal-poor solar-type star of magnitude V=11.9. We present in Section 2 the WASP discovery photometry plus high precision followup transit photometry and radial velocity measurements confirming the planetary nature of WASP-6b and including the observation of a spectroscopic transit. Section 3 presents the determination of the host star parameters. Our determination of the system parameters is presented in Section 4. These parameters are discussed in Section 5. 

\section{Observations}

\subsection{WASP photometry}

The host star 1SWASP J231237.75-224026.1 (= USNO-B1.0 0673-1077008 =  2MASS 23123773-2240261; hereafter WASP-6) was observed by WASP-South during the 2006 and 2007 observing seasons, covering the intervals 2006 May 07 to   2006 November 12 and 2007 July 05 to   2007 November 13 respectively. The 9630 pipeline-processed photometric measurements were de-trended and searched for transits using the methods described in Collier Cameron et al. (2006). The selection process (Collier Cameron et al. 2007) elected WASP-6 as a high priority candidate presenting a periodic transit-like signature with a period of 3.361 days. A total of 18 transits are observed in the data. Figure 1 presents the WASP photometry folded with the best-fit period.

\begin{figure}
\label{fig:a}
\centering                     
\includegraphics[width=9cm]{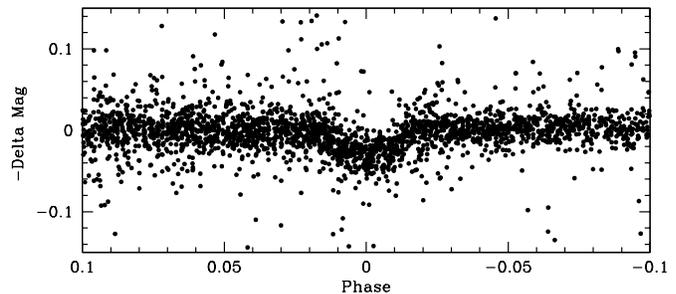}
\caption{WASP photometry of WASP-6 phase-folded with the best-fit period from the transit search algorithm presented in Collier Cameron et al. (2006).}
\end{figure}

\subsection{High-S/N transit photometry}

Followup transit photometry  was obtained on 2007 October 13 using the 2048$\times$2048 pixel$^2$ camera HawkCam2 (Wilson et al. 2008, Anderson et al. 2008) on the 2.0-m Faulkes Telescope South (FTS) at Siding Spring Observatory . The camera has a scale of 0.135 arcseconds/pixel and a field of view of $\sim 4.6\times4.6$ arcminutes$^2$.  We observed the target field using the SDSS $i$' band in the 2$\times$2 bin mode to improve the duty cycle. We acquired 247 frames of 60 sec exposure during the run. The telescope was sufficiently defocussed to keep the stellar flux within the linear range of the CCD. The images were bias subtracted and flat-field corrected with a master bias and twilight flat field images using {\tt IRAF}\footnote{ {\tt IRAF} is distributed by the National Optical Astronomy Observatory, which is operated by the Association of Universities for Research in Astronomy, Inc., under cooperative agreement with the National Science Foundation.}. {\tt DAOPHOT} aperture photometry (Stetson 1987) was performed around the target and comparison stars. We substracted a linear fit from the differential magnitudes as a function of airmass to correct for the different colour dependance of the  extinction for the target and comparison stars. The linear fit was calculated from the out-of-transit (OOT) data and applied to all the data. The corresponding fluxes were then normalized using the OOT part of the photometry. We discarded the first 17 measurements because they were obtained during twilight. Fig. 2 shows the resulting lightcurve folded on the best-fit orbital period and the residuals obtained after removing the best-fit transit model (see Section 4). Their $rms$ is $1.67 \times 10^{-3}$. This can be compared to $9.54 \times 10^{-4}$, the mean theoretical error bar taking into account photon, read-out, scintillation and background noises.

\begin{figure}
\label{fig:b}
\centering                     
\includegraphics[width=8cm]{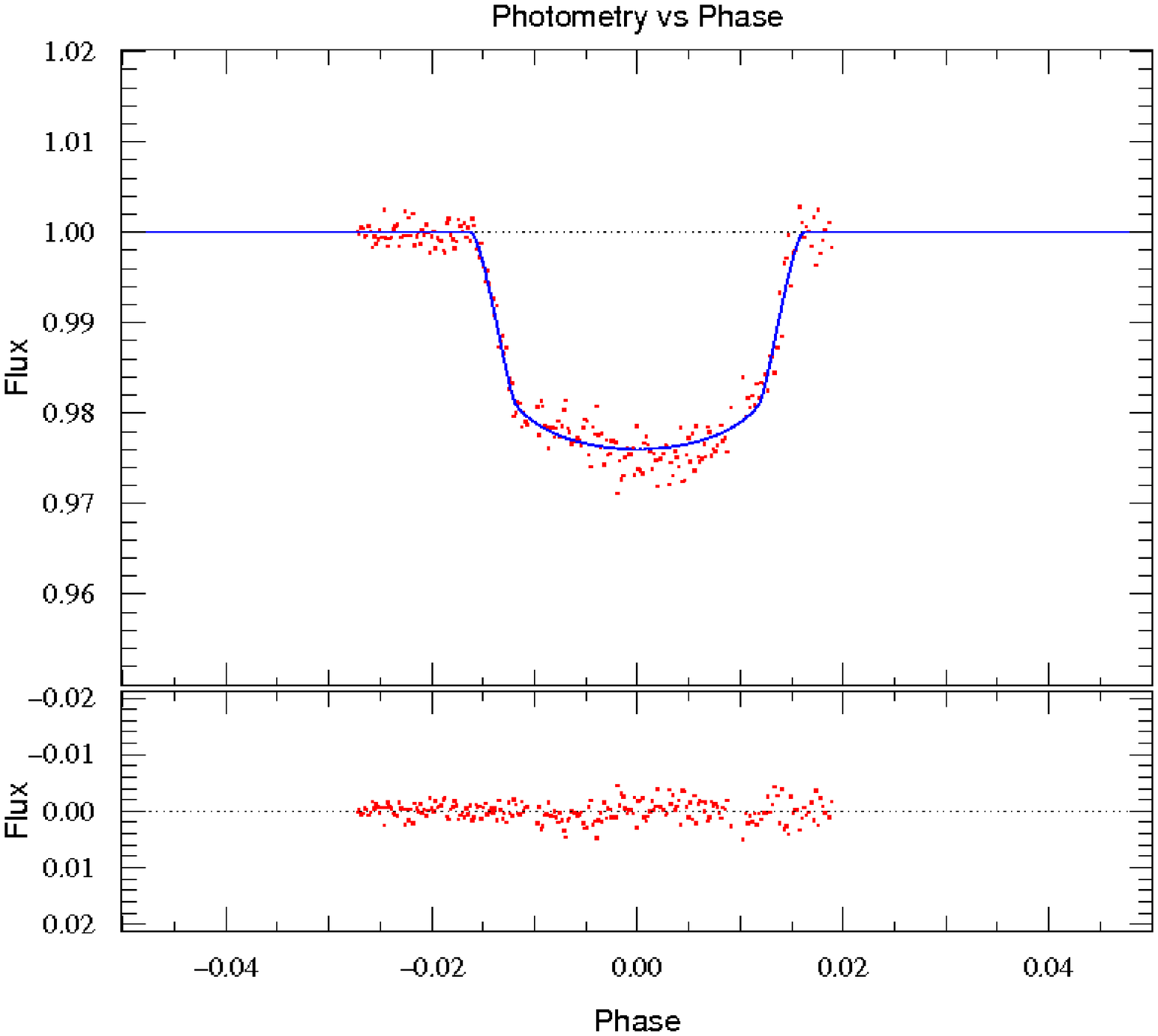}
\includegraphics[width=8cm]{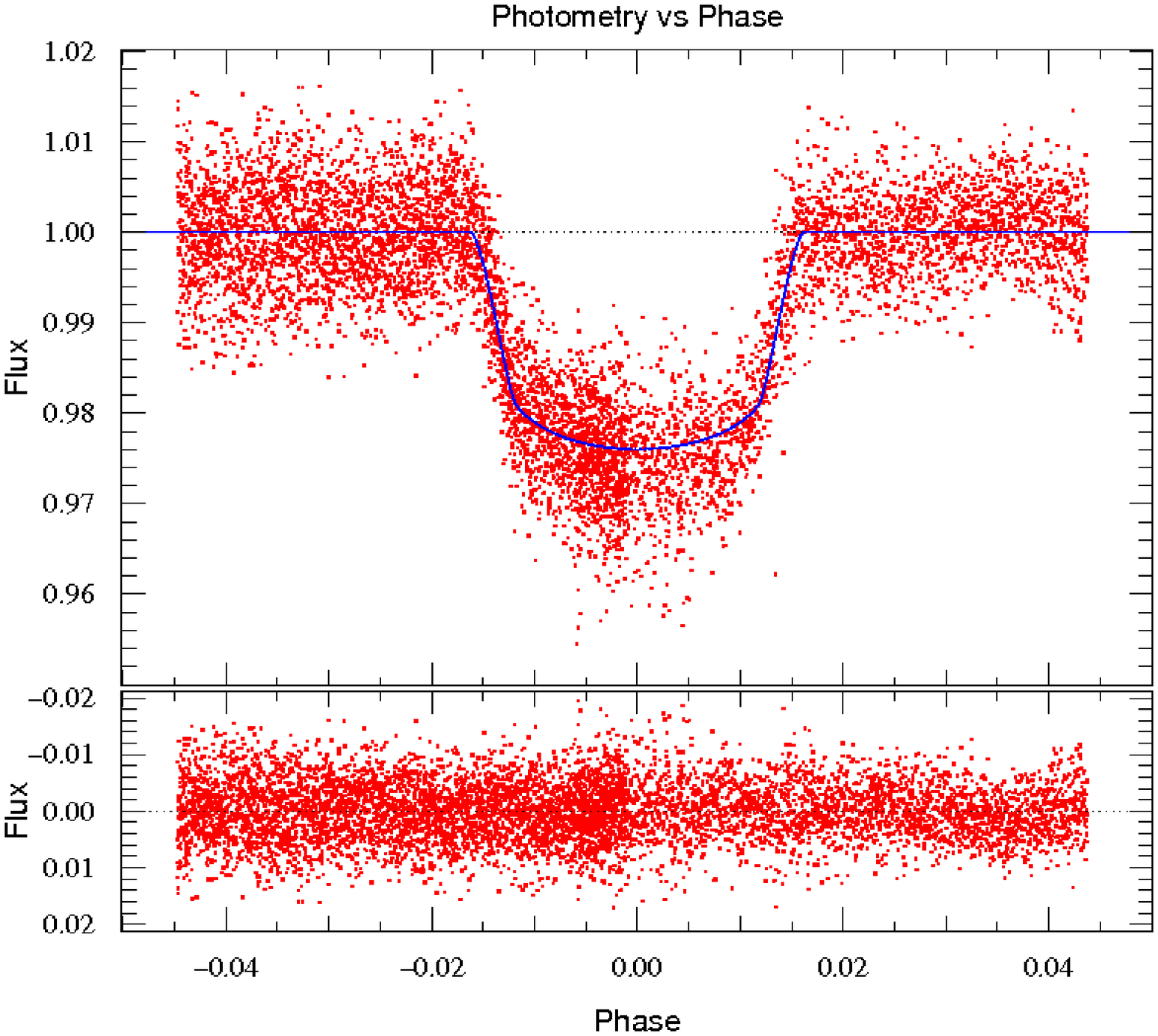}
\caption{FTS $i$'-band ($top$) and LT/RISE $V+R$ ($bottom$) transit photometry for WASP-6 and residuals after subtraction of the best-fit transit curve (superimposed in blue).}
\end{figure}

High precision transit observations of WASP-6 were also carried out using the 1024$\times$1024 pixel$^2$
thermoelectrically cooled frame transfer CCD camera RISE mounted on the 2-m Liverpool
Telescope (LT) in La Palma (Steele et al. 2008). The camera has a scale of 0.55 arcseconds/pixel and a total
field of view of $\sim 9.4\times9.4$ arcminutes$^2$. We observed the target field using a single
broad band $V+R$ filter in the 2$\times$2 bin mode. We acquired 4200 frames of 3 sec exposure on
the night of 2008 July 25 and 2880 frames of 5 sec exposure on the night of 2008 August 11. The telescope
was, here too,  defocussed. A similar reduction procedure as for the FTS photometry was used. The resulting normalized light curves of WASP-6 folded with the best-fit orbital period are shown in Fig. 2.  The $rms$ of the residuals is respectively 0.54 \% and 0.5 \% for the first and second run, while their mean theoretical error bar are 0.51 \% and 0.40 \%.

\subsection{Spectroscopy}

As soon as WASP-6 was identified as a high priority target, spectroscopic measurements were obtained using the CORALIE spectrograph mounted on the Euler Swiss telescope (La Silla, Chile) to confirm the planetary nature of the eclipsing body and measure its mass. WASP-6 was observed from 2007 September 16 to 2007 October 26 and from 2008 September 11 to 2008 September 25. Radial velocities (RV)  were computed by weighted cross-correlation (Baranne et al. 1996; Pepe et al. 2005) with a numerical G2-spectral template. RV variations of semi-amplitude $\sim$ 75 m~s$^{-1}$ were detected consistent with a planetary-mass companion whose period closely matches that from the WASP transit detections. 

44 additional spectroscopic measurements were obtained with the HARPS spectrograph (Mayor et al. 2003) based on the 3.6-m ESO telescope (La Silla, Chile) in the context of the programs 082.C-0040(E) and 082.C-0608(E). These programs aim to improve the characterization of WASP transiting planets. As CORALIE, HARPS is a cross-dispersed, fiber-fed, echelle spectrograph dedicated to high-precision Doppler measurements. HARPS data were reduced with a pipeline very similar to the CORALIE one. In addition to several measurements covering the whole orbital phase, high-cadence measurements of a spectroscopic transit were obtained with HARPS on 2008 October 08 in order to determine the sky-projected angle between the planetary orbital axis and the stellar rotation axis and included two points taken the night before, a point as far as possible from the transit on the transit night and a point the night after. This strategy aims to determine the systematic RV with greater accuracy than if the RM effect was taken on its own, assuming that stellar activity is the same over the three nights.

Our RV measurements are listed in Table 1 (CORALIE) and Table 2 (HARPS)  and are shown phase-folded and over-plotted with the best-fitting orbital+RM model in Fig. 3. 

To exclude that the RV signal shown in Fig. 3 is due to spectral line distortions cause by a blended eclipsing binary, the CORALIE and HARPS cross-correlation functions were analyzed using the line-bisector technique described in Queloz et al. (2001). No evidence for a correlation between the bisector spans and the RV variations was found (Fig. 4). The most likely cause for the periodic signal observed in photometry and RV measurements and for the RM effect observed on 2008 October 08 is thus the presence of a giant planet transiting the star WASP-6 every 3.36 days.

\begin{figure}
\label{fig:c}
\centering                     
\includegraphics[width=9cm]{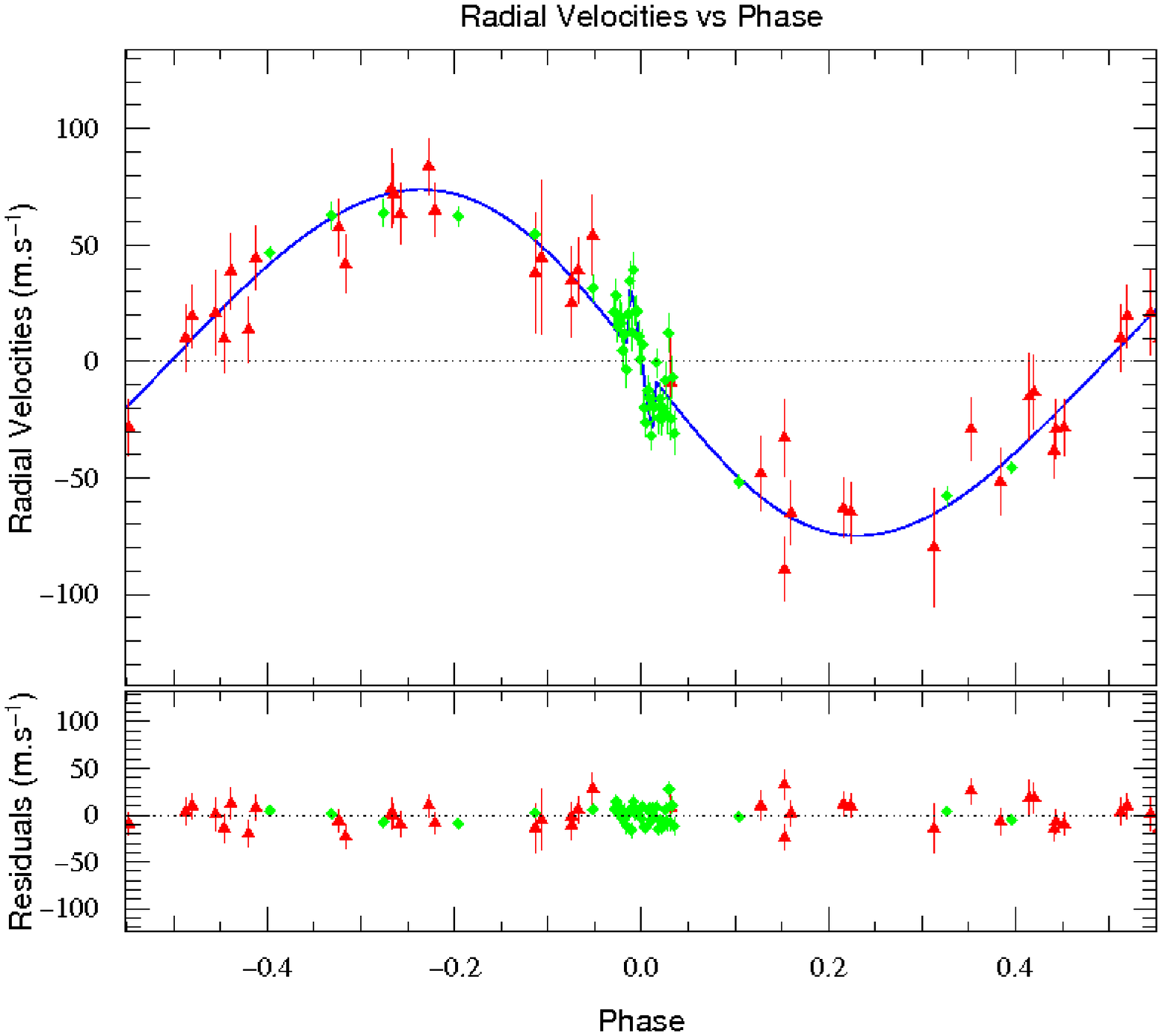}
\includegraphics[width=8.5cm]{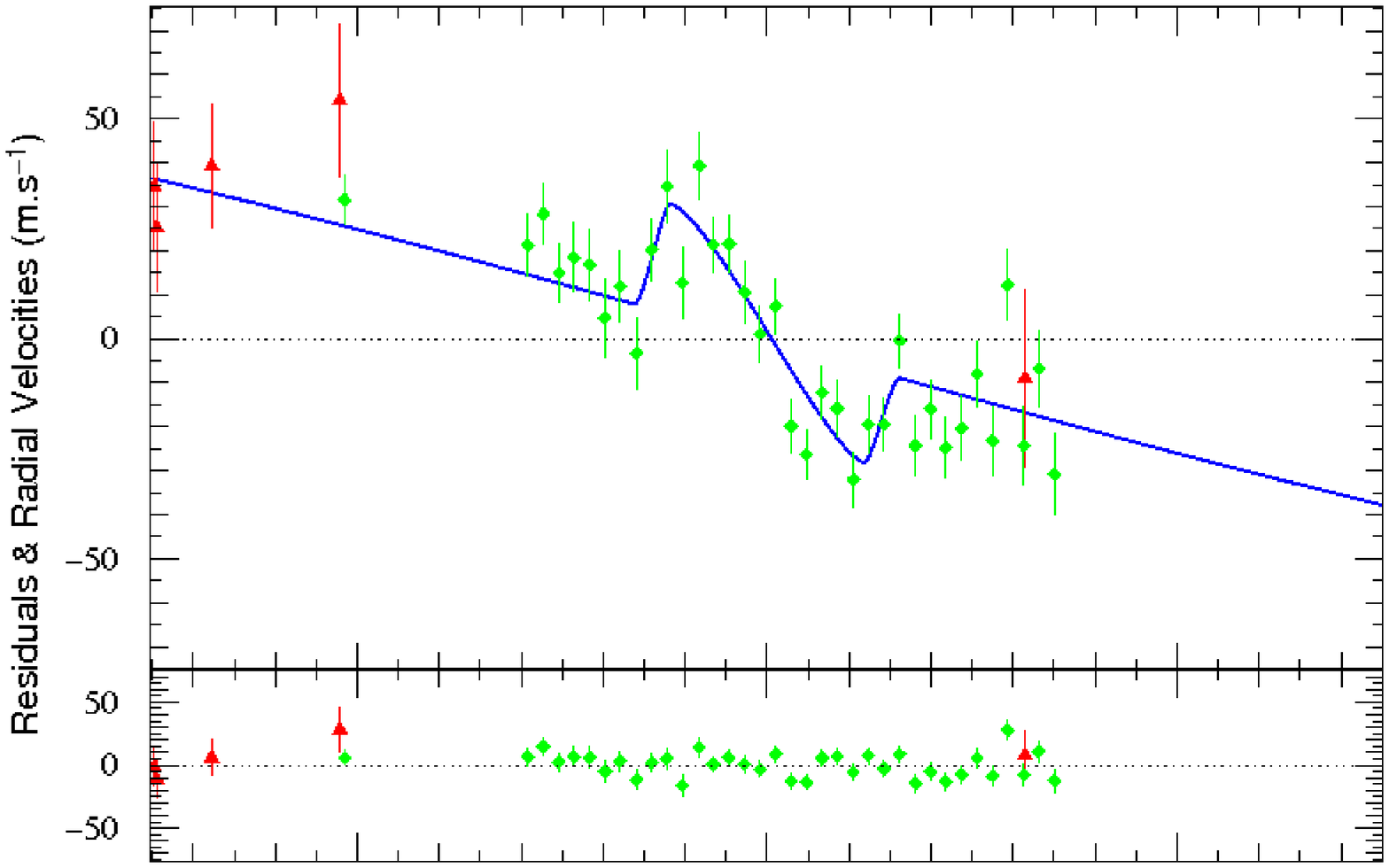}
\caption{$Top$: The RV measurements of WASP-6 obtained with CORALIE (red triangles) and HARPS (green squares). The systematic velocity has been subtracted. The solid line is the MCMC solution (see Section 4); it includes
the RM effect. $Bottom$: zoom on the transit phase showing the RM effect.}
\end{figure}

\begin{figure}
\label{fig:d}
\centering                     
\includegraphics[width=9cm]{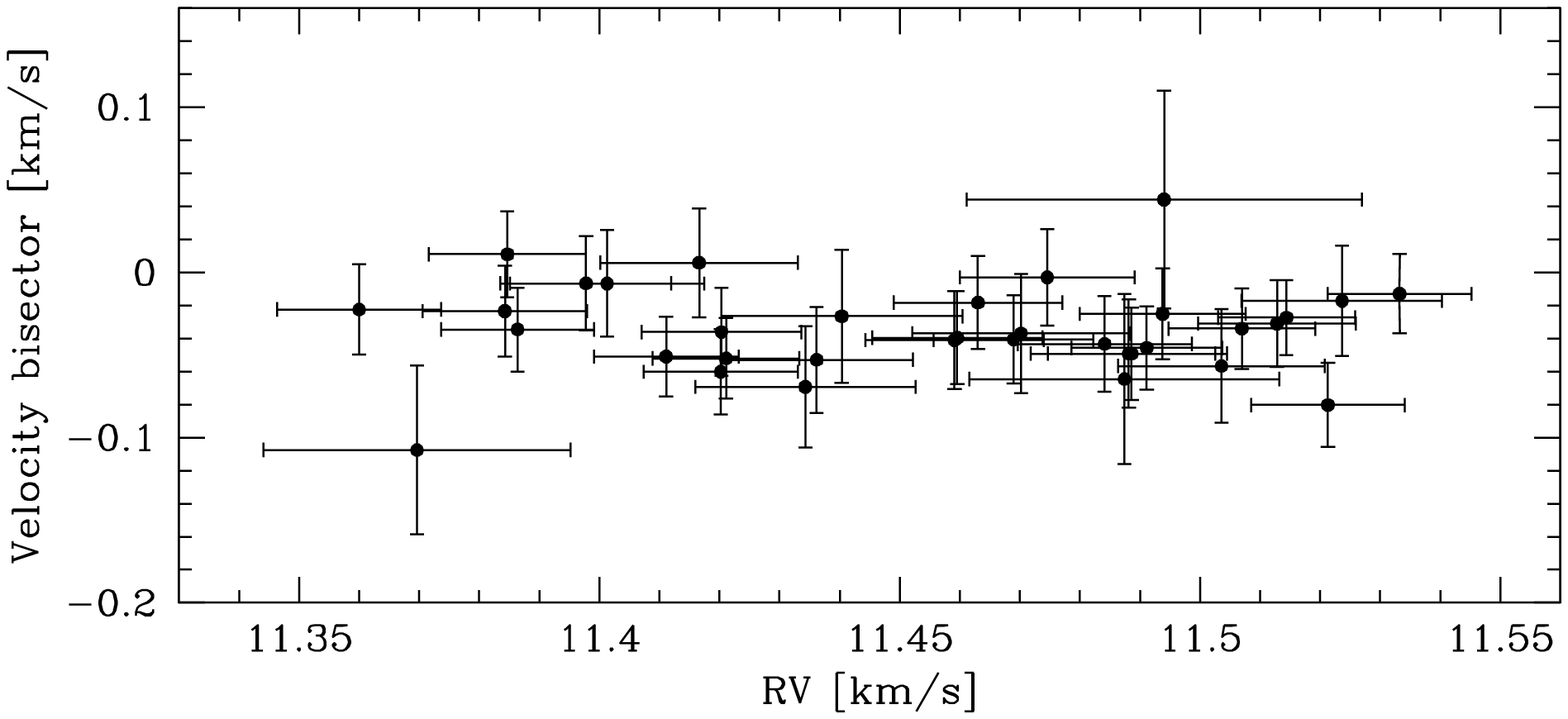}
\includegraphics[width=9cm]{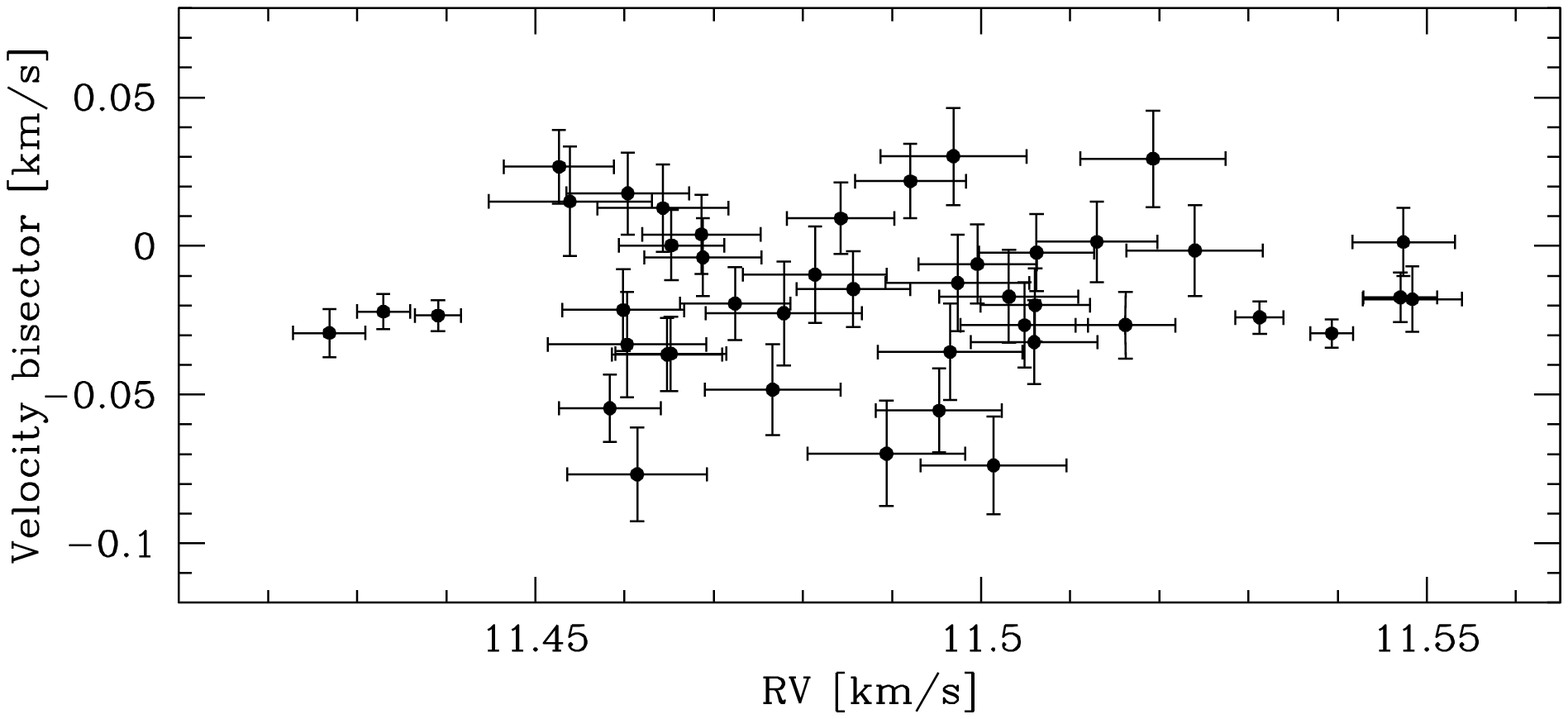}
\caption{Bisector versus RV measured from all the observed CORALIE ($top$) and HARPS ($bottom$) spectra. We adopt uncertainties of twice the RV uncertainty for all bisector measurements. There is no correlation between these two parameters indicating the RV variations are not caused by stellar activity or line-of-sight binarity.}
\end{figure}

\begin{table}
\begin{center}
\begin{tabular}{cccc}
\hline
BJD-2,400,000 & RV & $\sigma_{RV}$ & BS\\
(days) & (km~s$^{-1}$) & (km~s$^{-1}$) & (km~s$^{-1}$)\\
\hline
54359.687716 & 11.48815 & 0.01636 & 0.01618 \\
54362.573582	& 11.43618 & 0.01602 & -0.05289 \\
54364.628975	& 11.44037 & 0.02006 & -0.02655 \\
54365.707645	& 11.42032 & 0.01334 & -0.03610 \\
54372.735312	& 11.42022 & 0.01288 & -0.06013 \\
54377.716124	& 11.48411 & 0.01452 & -0.04330 \\
54377.739618	& 11.48861 & 0.01393 & -0.04916 \\
54378.693586	& 11.38638 & 0.01271 & -0.03449 \\
54378.719349	& 11.38469& 0.01306 &  0.01096 \\
54379.690005	& 11.45960 & 0.01414 & -0.03955 \\
54379.713452	& 11.46897& 0.01333 & -0.04065 \\
54380.563292	& 11.53322 & 0.01202 & -0.01290 \\
54380.586751	& 11.51439 & 0.01142 & -0.02735 \\
54382.617808	& 11.39778 & 0.01422 & -0.00658 \\
54383.601838	& 11.50697 & 0.01225 & -0.03398 \\
54383.625413	& 11.49114 & 0.01253 & -0.04571 \\
54385.740623	& 11.36965 & 0.02555 & -0.10750 \\
54386.638647	& 11.46303 & 0.01407 & -0.01829 \\
54386.664860	& 11.49377 & 0.01384 & -0.02509 \\
54387.668883	& 11.48742 & 0.02578 & -0.06456 \\
54387.690688	& 11.49402 & 0.03290 &  0.04418 \\ 
54390.521372	& 11.52130 & 0.01273 & -0.08028 \\
54390.544901	& 11.51282 & 0.01314 & -0.03094 \\
54398.647981	& 11.36002 & 0.01361 & -0.02246 \\
54398.671428	& 11.38428 & 0.01373 & -0.02346 \\
54399.616495	& 11.41113 & 0.01204 & -0.05097 \\
54399.652117	& 11.42110 & 0.01225 & -0.05205 \\
54720.614742	& 11.50358 & 0.01721 & -0.05665 \\
54722.651535	& 11.45913 & 0.01483 & -0.04094 \\
54724.579899	& 11.40131 & 0.01616 & -0.00673 \\
54725.545387	& 11.43430 & 0.01835 & -0.06935 \\
54726.613463	& 11.52363& 0.01670 & -0.01711 \\
54730.622609	& 11.47459 & 0.01459 & -0.00302 \\
54732.703073	& 11.47021 & 0.01808 & -0.03689 \\
54734.747409	& 11.41662 & 0.01649 &  0.00588 \\
\hline
\end{tabular}
\caption{CORALIE radial velocity measurements for WASP-6 (BS = bisector spans).}
\end{center}
\label{wasp6-params}
\end{table}

\begin{table}
\begin{center}
\begin{tabular}{cccc}
\hline
BJD-2,400,000 & RV & $\sigma_{RV}$ & BS\\
(days) & (km~s$^{-1}$) & (km~s$^{-1}$) & (km~s$^{-1}$)\\
\hline
54746.564605  &  11.54738   &     0.00576 & 0.00133 \\
54746.750752   & 11.54840   &     0.00555 &-0.01787 \\
54747.504842   & 11.51620   &     0.00559 &-0.02660 \\
54747.579780   & 11.50597   &     0.00711 &-0.03234 \\
54747.586192   & 11.51299   &     0.00677 & 0.00139 \\
54747.592603   & 11.49959   &     0.00665 &-0.00608 \\
54747.598784   & 11.50313   &     0.00781 & -0.01701\\
54747.605195   & 11.50141   &     0.00819 & -0.07376\\
54747.611549   & 11.48940   &     0.00885 & -0.06980\\
54747.617718   & 11.49653   &     0.00811 & -0.03563\\
54747.624372   & 11.48137   &     0.00808 & -0.00962\\
54747.630668   & 11.50486   &     0.00716 & -0.02661\\
54747.636837   & 11.51929   &     0.00812 & 0.02936 \\
54747.643307   & 11.49737   &     0.00808 & -0.01246\\
54747.650089   & 11.52399   &     0.00766 & -0.00152\\
54747.655956   & 11.50607   &     0.00612 & -0.01990\\
54747.662310   & 11.50622   &     0.00648 & -0.00217\\
54747.668664   & 11.49528   &     0.00706 & -0.05534\\
54747.675076   & 11.48567   &     0.00634 & -0.01449\\
54747.681372   & 11.49207   &     0.00624 & 0.02179 \\
54747.687841   & 11.46478   &     0.00616 & -0.03649\\
54747.694137   & 11.45835   &     0.00571 & -0.05455\\
54747.700491   & 11.47240   &     0.00618 & -0.01938\\
54747.706729   & 11.46878   &     0.00655 & -0.00382\\
54747.713256   & 11.45266   &     0.00619 & 0.02667 \\
54747.719494   & 11.46519   &     0.00625 & -0.03618\\
54747.725906   & 11.46526   &     0.00591 & 0.00022 \\
54747.732260   & 11.48427   &     0.00603 & 0.00930 \\
54747.738672   & 11.46037   &     0.00689 & 0.01768 \\
54747.744968   & 11.46863   &     0.00668 & 0.00392 \\
54747.751264   & 11.45986   &     0.00685 & -0.02154\\
54747.757617   & 11.46432   &     0.00734 & 0.01274 \\
54747.764029   & 11.47662   &     0.00759 & -0.04832\\
54747.770383   & 11.46143   &     0.00783 & -0.07680\\
54747.776621   & 11.49690   &     0.00818 & 0.03016 \\
54747.783206   & 11.46028   &     0.00886 & -0.03313 \\
54747.789329   & 11.47787   &     0.00875 & -0.02263 \\
54747.795798   & 11.45389   &     0.00916 & 0.01502 \\
54748.775349   & 11.42689   &     0.00406 & -0.02922\\
54749.703663   & 11.53125   &     0.00270 & -0.02405\\
54750.655365   & 11.53933   &     0.00240 & -0.02941\\
54753.741910   & 11.54705   &     0.00414 & -0.01725\\
54754.750340   & 11.43295   &     0.00299 & -0.02203\\
54755.730495   & 11.43908   &     0.00261 & -0.02337\\
\hline
\end{tabular}
\caption{HARPS radial velocity measurements for WASP-6 (BS = bisector spans).}
\end{center}
\label{wasp7-params}
\end{table}

\section{WASP-6 Stellar Parameters}

The individual CORALIE and HARPS spectra are relatively low signal-to-noise,
but when co-added into 0.01\AA\ steps they give a S/N of in excess of 100:1
which is suitable for a photospheric analysis of WASP-6. The standard pipeline
reduction products were used in the analysis.

The analysis was performed using the {\sc uclsyn} spectral synthesis package
(Smith 1992; Smalley et al. 2001) and {\sc atlas9} models without convective
overshooting (Castelli, Gratton \& Kurucz 1997). The \halpha\ line were used to
determine the effective temperature (\teff), while the Na {\sc i} D and Mg {\sc
i} b lines were used as surface gravity (\logg) diagnostics.  The parameters
obtained from the analysis are listed in Table~\ref{wasp6-params}.

The equivalent widths of several clean and unblended lines were measured.
Atomic line data was mainly taken from the Kurucz \& Bell (1995) compilation,
but with updated van der Waals broadening coefficients for lines in Barklem et
al. (2000) and $\log gf$ values from Gonzalez \& Laws (2000), Gonzalez et al.
(2001) or Santos et al. (2004). A value for microturbulence (\mictrb) was
determined from Fe~{\sc i} using Magain's (1984) method. The ionization balance
between Fe~{\sc i} and Fe~{\sc ii} and the null-dependence of abundance on
excitation potential were used as an additional the \teff\ and \logg\
diagnostics (Smalley 2005).

We have determined the elemental abundances of several elements (listed in
Table~\ref{wasp6-params}) from their measured equivalent widths. The quoted error
estimates include that given by the uncertainties in \teff, \logg\ and \mictrb,
as well as the scatter due to measurement and atomic data uncertainties. In our
spectra the Li {\sc i} 6708\AA\ line is not detected (EW $<$ 2m\AA), allowing
us to derive an upper-limit on the Lithium abundance of log n(Li/H) + 12 $<$
0.5. The lack of lithium implies an age in excess of $\sim$3~Gyr (Sestito \&
Randich 2005).

Projected stellar rotation velocity ($V_{rot} \sin I$) was determined by fitting the profiles of
several unblended Fe~{\sc i} lines in the HARPS spectra. We used 
a value for macroturbulence (\mactrb, see Gray 2008) of 2~\kms\ and an
instrumental FWHM of 0.060 $\pm$ 0.005 \AA, determined from the telluric lines
around 6300\AA. A best fitting value of $V_{rot}\sin I$ = 1.4 $\pm$ 1.0~\kms\ was
obtained. If, however, mactroturbulence is lower, then higher rotation values
are found, with $V_{rot} \sin I$ = 3.0 $\pm$ 0.5~\kms\ obtained for \mactrb\ = 0~\kms.
If, on the other hand, \mactrb\ is slightly higher than 2~\kms, then it is
possible that $V_{rot} \sin I$  is close to, or even, zero.

In addition to the spectral analysis, we have also used broad-band photometry
from TYCHO-2, USNO-B1.0 R-mag., CMC14 r', DENIS and 2MASS to estimate the total
observed bolometric flux. The Infrared Flux Method (Blackwell \& Shallis 1977)
was then used with 2MASS magnitudes to determine \teff\ and stellar angular
diameter ($\theta$). This gives \teff = 5470 $\pm$ 130~K, which is in close
agreement with that obtained from the spectroscopic analysis and implies a
spectral type of G8V (Gray 2008).

\begin{table}[h]
\begin{center}
\begin{tabular}{ccccccc}
\hline
Parameter  &&&&&& Value\\
\hline
R.A. (J2000)  &&&&&& 23$^h$12$^m$37.74$^s$\\
Dec (J2000) &&&&&& -22$^{\circ}$40'26''.2\\
V &&&&&& 11.9\\
\\
\teff     &&&&&& 5450 $\pm$ 100 K \\
\logg       &&&&&&4.6 $\pm$ 0.2 \\
\mictrb    & &&&&&1.0 $\pm$ 0.2 \kms \\
$V_{rot} \sin I$     & &&&&&1.4 $\pm$ 1.0 \kms \\
\\
{[Na/H]}   &&&&&&$-$0.17 $\pm$ 0.06 \\
{[Mg/H]}   &&&&&&$-$0.13 $\pm$ 0.07 \\
{[Al/H]}   &&&&&&$-$0.15 $\pm$ 0.10 \\
{[Si/H]}   &&&&&&$-$0.12 $\pm$ 0.08 \\
{[Ca/H]}   &&&&&&$-$0.09 $\pm$ 0.10 \\
{[Sc/H]}   &&&&&&$-$0.22 $\pm$ 0.15 \\
{[Ti/H]}   &&&&&&$-$0.05 $\pm$ 0.09 \\
{[V/H]}    &&&&&&$-$0.02 $\pm$ 0.08 \\
{[Cr/H]}   &&&&&&$-$0.17 $\pm$ 0.09 \\
{[Mn/H]}   &&&&&&$-$0.20 $\pm$ 0.13 \\
{[Fe/H]}   &&&&&&$-$0.20 $\pm$ 0.09 \\
{[Co/H]}   &&&&&&$-$0.16 $\pm$ 0.14 \\
{[Ni/H]}   &&&&&&$-$0.21 $\pm$ 0.08 \\
\\
$\log N(Li)$&&&&&& $<$ 0.5 \\
\teff(IRFM)&&&&&&5470 $\pm$ 130 K\\
$\theta$(IRFM)&&&&&& 0.037 $\pm$ 0.002 mas \\
\hline
\end{tabular}
\caption{Stellar parameters for WASP-6.}\label{wasp6-params}
\end{center}
\end{table}

\section{Derivation of the system parameters}

We derived stellar and planetary parameters for the system by fitting simultaneously the WASP,  FTS and LT/RISE photometry with the CORALIE and HARPS RVs. These data were used as input into the Markov Chain Monte Carlo (MCMC; Ford 2006) code described in Gillon et al. (2008). MCMC is a Bayesian inference method based on stochastic simulations and provides the $a$ $posteriori$ probability distribution of adjusted parameters for a given model. Here the model is based on a star and a transiting planet on a keplerian orbit about their center of mass. Specifically, we used the photometric transit model of Mandel \& Agol (2002) and the spectroscopic transit model of Gim\'enez (2006) in addition to a classical Keplerian model for the orbital part of the RV variations.  To model the transit lightcurves, a quadratic limb darkening law was assumed, with coefficients interpolated from Claret's tables (2000; 2004) for the appropriate photometric filters.  For the RISE broad band filter, the average from V and R bands was taken to be our theoretical limb darkening parameters.

We used 16 jump parameters  in our MCMC simulations: the orbital period $P$, the time of minimum light $T_0$, the transit depth $D$, the total transit width $W$, the impact parameter $b$, the stellar mass $M_\ast$, the orbital RV semi-amplitude $K$, a systematic radial velocity $\gamma$ for each spectroscopic instrument (HARPS and CORALIE),  the parameters $e\cos \omega$ and $e\sin\omega$ where $e$ is the orbital eccentricity and $\omega$ is the argument of periastron, the products $V_{rot} \sin I \cos \beta$ and $V_{rot} \sin I  \sin \beta$ where $V_{rot} \sin I$ is the projected stellar rotational velocity and  $  \beta$  is the spin-orbit angle (see Gim\'enez 2006), and a normalization factor for each of the 4 light curves (assuming the same normalization for the whole SW photometry).  As explained in the now abundant literature on the application of MCMC to perform Bayesian inference for transiting planets (see e.g. Collier Cameron et al. 2007 and references therein), each MCMC simulation is composed of a large number of consecutive steps for which the jump parameters are randomly modified or not depending of the result of a test on the merit function ($MF$). The $MF$ used here is the sum of the $\chi^2$ for all the data with respect to the models added to a Bayesian prior on $V_{rot} \sin I$ and $M_\ast$ representing our constraints on these parameters from spectroscopy: 

\begin{equation}\label{eq:c}
MF =  \chi^2 + \frac{(V_{rot} \sin I - (V_{rot} \sin I)_0)^2}{\sigma_{V_{rot} \sin I}^2} + \frac{(M_\ast - (M_\ast)_0)^2}{\sigma_{M_\ast}^2} \end{equation}\noindent

where $(V_{rot} \sin I)_0$ = 1.4 \kms, $\sigma_{V_{rot} \sin I}$ = 1 \kms, $M_\ast$ = 0.87 and $\sigma_{M_\ast}$ = 0.08. These last two values were obtained  by interpolation of Girardi stellar evolution models (Girardi et al. 2000) in order to find the mass and age that best match the spectroscopic parameters. We notice that our data do not constrain strongly $M_\ast$ and that it is a free parameter under the control of a Bayesian prior in our simulations only to propagate its uncertainty to the other physical parameters. 

A first MCMC run was performed and led to a refined value for the stellar density. We
 converted it to $R_\ast/M_\ast^{1/3}$ in solar units, and compared this property and the stellar temperature to the
Girardi models interpolated at -0.2 metallicity.  The quantity, $R_\ast/M_\ast^{1/3}$,
depends only on the observed transit properties (duration, depth, impact parameter, and orbital period) and is independent of the measured temperature. We generated the same property from the mass and log $g$ values in the models, and then interpolated the models in the $R/M^{1/3}$-\teff $ $ plane to determine
a mass and age for WASP-6.  We interpolated linearly along two consecutive
mass tracks to generate an equal number of age points between the
zero-age main sequence and the evolutionary state  where the star reaches the 
end of core hydrogen burning.  We then interpolated
between the mass tracks along equivalent evolutionary points to find the
mass and age from the models that best match the stellar density derived
from the MCMC and the effective temperature.  In this way,
we obtained a value for the stellar mass of, $M_\ast =0.83^{+0.07}_{-0.09}$ $M_{\odot}$ and 
 a derived age for the system of $11 \pm 7$~Gyr.

The best-fitting model found in the first MCMC run was used to estimate the level of correlated noise in each photometric time-series and a jitter noise in the RV time series. For each photometric time-series,  the red noise was estimated as described in Gillon et al. (2006), by comparing the $rms$ of the unbinned and binned residuals.  We used a bin size corresponding to a duration of 25 minutes, similar to the timescale of the ingress/egress of the transit. For the SW data, the red noise was estimated to be negligible when compared to the theoretical error bar of the measurements and it was thus neglected. The deduced red noise values (Table 4) were added quadratically to the theoretical uncertainties of each corresponding time-series. No jitter is detected in the CORALIE data. For the HARPS data, a significant jitter is obtained, but it seems to be originating mostly from the residuals of the RM effect and is probably more due to lower-than-usual $S/N$ on the spectra and a worsening of airmass (reaching 1.8 at the end of the sequence) than stellar activity. For this reason, no jitter noise was added to the RV uncertainties.


Using the updated value of the stellar mass as initial value, a  second  MCMC run was then performed. This chain allowed a large safety burn-in period of discarded 50,000 steps followed by a simulation of 500,000 steps allowing a robust determination of the $a$ $posteriori$ probability distributions for the jumped parameters. The parameters set (jump + deduced parameters) corresponding to the lowest $MF$ was considered as the best solution, and for each parameter upper and lower 1-$\sigma$ error bars were obtained from respectively the 68.3\% larger and smaller values. Best-fitting jump + physical parameters are shown in Table 5. The reduced $\chi^2$ of the best-fitting solution is 0.86. 

\begin{table*}[t]
\begin{center}
\begin{tabular}{ccc}
\hline
\hline \noalign {\smallskip}
Photometric time-series & Red noise [ppm] & \\ \noalign {\smallskip}
\hline
FTS & 545 & \\
RISE-1 &  317 &  \\
RISE-2 &  770 & \\
\hline \noalign {\smallskip}
RV time-series & Jitter [\ms] & systematic RV [\kms] \\ \noalign {\smallskip}
 \hline 
 CORALIE  & 0 & 11.449\\
HARPS  & 6.4 & 11.485 \\
\hline
\end{tabular}
\caption{Deduced values for the photometric red noise ($top$), and the RV jitter and systematic velocities ($bottom$).}\label{wasp6-jitter}
\end{center}
\end{table*}

\begin{table*}
\begin{center}
\label{tab:params}
\begin{tabular}{lcccl}
\hline
Parameter  & Value & Units \\
\hline  \noalign {\smallskip}
Transit epoch  $ T_0  $ & $ 2454596.43267^{+ 0.00015}_{- 0.00010} $  & HJD \\  \noalign {\smallskip}
Orbital period  $ P  $ & $ 3.3610060^{+ 0.0000022 }_{- 0.0000035 } $ & days \\  \noalign {\smallskip}
Planet/star area ratio  $ (R_p/R_s)^2 $ & $ 0.02092 ^{+ 0.00019 }_{- 0.00025 } $ &  \\  \noalign {\smallskip}
Transit duration  $ t_T $ & $ 0.10860^{+ 0.00073}_{- 0.00067} $  & days \\  \noalign {\smallskip}
Impact parameter  $ b $ & $ 0.26^{+ 0.07}_{- 0.11} $  & $R_*$ \\  \noalign {\smallskip}
RV semi-amplitude  $K$ & $74.3^{+1.7}_{-1.4}$ & \ms \\  \noalign {\smallskip}
$e \cos\omega$ & $-0.007^{+0.011}_{-0.008}$ & \\  \noalign {\smallskip}
$e \sin\omega$  & $0.054^{+0.018}_{-0.017}$ & \\  \noalign {\smallskip}
$V_{rot} \sin I \cos\beta$  &  $1.57^{+0.28}_{-0.10}$& \\  \noalign {\smallskip}
$V_{rot} \sin I  \sin\beta$  &  $0.32^{+0.49}_{-0.50}$ & \\  \noalign {\smallskip}
      &      &  \\
Orbital semi-major axis $ a $ & $ 0.0421 ^{+ 0.0008}_{- 0.0013} $  & AU \\  \noalign {\smallskip}
Orbital inclination  $ i $ & $ 88.47^{+ 0.65 }_{- 0.47} $ & degrees \\  \noalign {\smallskip}
Orbital eccentricity $ e $ & $ 0.054^{+ 0.018 }_{- 0.015} $ &  \\  \noalign {\smallskip}
Argument of periastron  $ \omega $ & $1.70^{+0.12}_{-0.23}$  & rad  \\  \noalign {\smallskip}
Spin-orbit angle $ \beta$    &   $0.20^{+0.25}_{-0.32}$  & rad  \\  \noalign {\smallskip}
& & \\
 \noalign {\smallskip}
Stellar mass  $ M_\ast $ & $ 0.880^{+ 0.050 }_{- 0.080}$ & $M_\odot$ \\  \noalign {\smallskip}
Stellar radius  $ R_\ast $ & $ 0.870^{+ 0.025}_{- 0.036} $ & $R_\odot$ \\  \noalign {\smallskip}
Stellar surface gravity  $ \log g_* $ & $4.50 \pm 0.06 $ & [cgs] \\  \noalign {\smallskip}
Stellar density  $ \rho_\ast $ & $ 1.34^{+ 0.11}_{- 0.10} $ & $\rho_\odot$ \\  \noalign {\smallskip}
Projected rotational velocity $V_{rot} \sin I$ & $1.60^{+0.27}_{-0.17}$ & \kms \\
\noalign {\smallskip}    &      &  \\
Planet radius  $ R_p $ & $ 1.224 ^{+ 0.051}_{- 0.052} $ & $R_J$ \\  \noalign {\smallskip}
Planet mass  $ M_p $ & $ 0.503 ^{+ 0.019}_{- 0.038} $  & $M_J$ \\  \noalign {\smallskip}
Planetary surface gravity  $ \log g_p $ & $  7.857 \pm  0.028$ & [cgs] \\  \noalign {\smallskip}Planet density  $ \rho_p $ & $0.27 \pm 0.05$&  $\rho_{J}$ \\  \noalign {\smallskip}
Planet temperature ($A=0, f=1/4$)   $ T_{\mbox{eff}} $ & $1194^{+58}_{-57}$ &  K \\  \noalign {\smallskip}

\hline\\
\end{tabular}
\caption[]{WASP-6 system parameters and 1-$\sigma$ error limits derived in this work.}
\end{center}
\end{table*}

\begin{figure}
\label{fig:e}
\centering                     
\includegraphics[width=9cm]{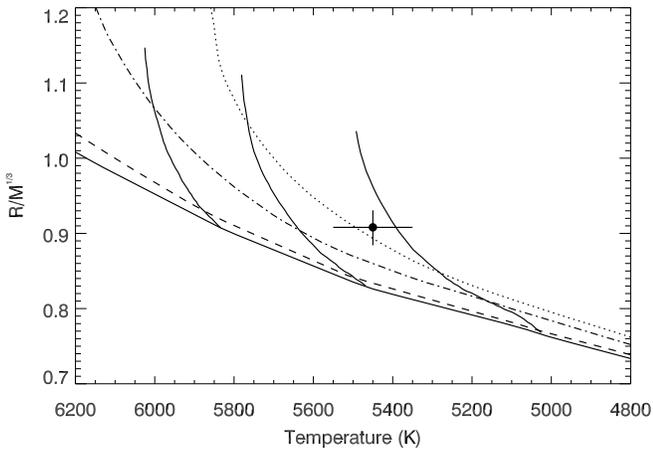}
\caption{$R/M^{1/3}$ (in solar units) versus effective temperature 
for WASP-6 compared to the theoretical stellar stellar evolutionary 
models of Girardi et al. (2000). The labeled mass tracks are for 0.8, 0.9 and 1.0 $M_\odot$ and the isochrones are 100 Myr (solid), 1 Gyr (dotted), 5 Gyr (dot-dashed), 10 Gyr (dotted).  We  have interpolated the tracks at -0.2 metallicity and have included the uncertainty on the metallicity (+-0.1) in the overall uncertainties on the mass and the age. According to
the models, the host star has an age of $11 \pm 7$ Gyr. }
\end{figure}

\section{Discussion}

The large radius of WASP-6b  ($\sim 1.2 R_J$) and the metal deficiency of its host star strengthen the existence of a correlation between the heavy-element content of giant planets and the stellar metallicity (Guillot et al. 2006; Burrows et al. 2007). With half of the mass of Jupiter and  a radius significantly larger, WASP-6b appears nevertheless too large for basic models of irradiated planets (Burrows et al. 2007a; Fortney et al. 2007), even if an absence of core is assumed. For instance, tables presented in Fortney et al. (2007) predict a maximum radius of $\sim 1.1 R_J$ for a 0.5 Jupiter-mass planet orbiting at 0.045 AU of a 4.5 Gyr solar-type star. WASP-6 is smaller, cooler and probably older than the Sun, so $1.2 R_J$ is clearly too large for these models.   In this context, it is worth noticing the non-null eccentricity that we infer for its orbit ($e =  0.054^{+ 0.018 }_{- 0.015} $). The fact that the planetary orbit is still not circularized despite the  large age of the system indicates that the tidal evolution of WASP-6b probably played an important role in its energy budget.  As outlined by Jackson et al. (2008b), tidal heating could have been large enough for many close-in planets to explain at least partially the large radius of some of them. To assess the past and future tidal evolution of  WASP-6b, we integrated the equations for $da/dt$ and $de/dt$ presented in Jackson et al. (2008a)  and computed at each step the tidal heating rate $H$ using the formula presented in Jackson et al. (2008b). We assumed values of $Q_p' = 10^{6.5}$ and $Q_\ast' = 10^{5.5}$ for respectively the planetary and stellar tidal dissipation parameters\footnote{We use here the same convention than Jackson et al. (2008a): the coefficients $Q_p'$ and $Q_\ast'$ used here are equal to the actual tidal dissipation parameters $Q_p$ and $Q_\ast$ multiplied by the ratio $3/2k$ where $k$ is the Love number. }. These values were found by Jackson et al. (2008a) to conciliate the eccentricity distribution of close-in planets before their tidal evolution to the one of the planets detected farer from their star. We also took into account the evolution of the stellar rotation period due to the tide raised by the planet  using (Goldreich \& Soter 1966): 

\begin{equation}\label{eq:d}
\frac{d\Omega_\ast}{dt} =  -sign(\Omega_\ast - n)  \frac{9}{4}  G \frac{R_\ast^3}{\alpha_\ast M_\ast Q_\ast'} \frac{M_p^2}{a^6}\textrm{,}
\end{equation}\noindent
where $G$ is the gravitational constant,  $n$ is the mean orbital motion, $\Omega_\ast$ is the stellar spin angular rate and $\alpha_\ast$ = $I_\ast/(M_\ast R_\ast^2)$ with $I_\ast$ being the moment of inertia though the spin axis of the star. For $\alpha_\ast$, we assumed a value of 0.07 (P\"atzold et al. 2004). To assess the reliability limits of the model,  we also computed the evolution of the total angular momentum of the system (assuming a negligible contribution of the planet rotation): 

\begin{equation}\label{eq:d}
L_{tot} =  \frac{M_\ast  M_p}{M_\ast + Mp} n a^2 \sqrt{1 - e^2} + \alpha_\ast M_\ast R_\ast^2 \Omega_\ast \textrm{,}
\end{equation}\noindent
 Neglecting the possible decrease due stellar wind (Dobbs-Dixon et al. 2004), $L_{tot}$ should be a conserved quantity during the whole tidal evolution of the system. 

Fig. 6 shows the obtained evolution for $a$, $e$, $H$, $L_{tot}$ and the orbital and stellar rotation period from 2 Gyr ago to  5 Gy in the future.  Interestingly, the model predicts (1) that the eccentricity and semi-major axis of WASP-6b were significantly larger in the past, (2) that the orbit will be fully circularized one Gy from now, and (3) that the planet will continue to slightly approach the star until finally reaching its Roche limit. This last results agrees well with the fact that the ratio $L_{tot}/L_c$, where $L_c$ is critical angular momentum (see Levrard et al. 2009), has a value of $\sim$ 0.6, implying that the system is tidally unstable and will ultimately merge. Levrard et al. (2009) showed that all the other transiting systems, except HAT-P-2, are in the same case.

Under this tidal evolution model, WASP-6b was brought to a distance $> 0.05$ AU of its host star in the very early life of the system, then its orbital evolution has been totally dominated by tides until now. This evolution does not consider the possible influence of one or more other planets able to pump the eccentricity of WASP-6b (Mardling 2007), but our RV data do not reveal the presence of another planet so it seems reasonable at this stage to assume that the orbital evolution of WASP-6b was not dominated by planet-planet interactions. The model assumes also a constant radius for the planet during the whole tidal evolution, which is not very likely (Liu et al. 2008). Furthermore, Fig. 6 shows that it does not conserve $L_{tot}$ for $e >\sim 0.3$ and  during the final runaway merging of the planet with the star. Considering as valid only the part of the tidal evolution for which $L_{tot}$ is conserved at the 1-\% level, we can nevertheless conclude from Fig. 6 that WASP-6b experienced 0.6 - 1.2 Gyr ago a large tidal heating rate of 5 -10  $\times 10^{19}$ W.  Such a large heating rate in the past should have modified drastically the thermal history of the planet and could have contributed significantly to the measured inflated radius.  
 
\begin{figure}\label{fig:f}
\centering                     
\includegraphics[width=9cm]{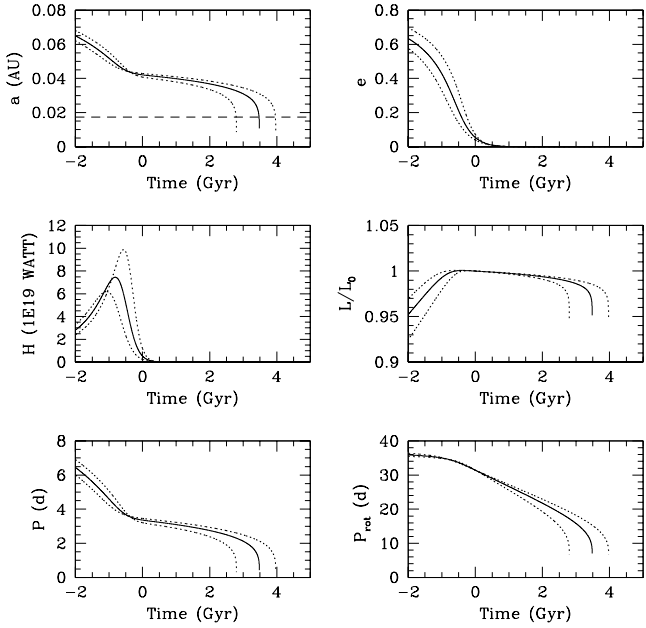}
\caption{Tidal evolution for WASP-6b computed using the method described in Jackson et al. (2008b) . {\it Top left}: evolution of the semi-major axis. The dashed line shows the Roche limit of the system. {\it Top right}: evolution of the eccentricity. {\it Middle left}: evolution of the tidal heating rate. {\it Middle right}: evolution of the  total angular momentum. {\it Bottom}: evolution of the orbital ({\it left}) and star rotational ({\it right})  period. For each parameter, the solid line shows the evolution computed with the best-fitting present eccentricity and semi-major axis while the dotted lines assume the maximum and minimum tidal heating consistent with their 1-$\sigma$ error bars.}
\end{figure} 
 
With a stellar irradiation $\sim$ 4.7 10$^8$ erg s$^{-1}$ cm$^{-2}$, WASP-6b belongs to the theoretical pL planetary class proposed by Fortney et al. (2008; see also Burrows et al. 2008). Under this theory, Ti and V-bearing compounds should mostly be condensed in the planetary atmosphere and secondary eclipse measurements at different wavelengths should not reveal any stratospheric thermal inversion. Such secondary eclipses observations would not only constraint atmospheric models of giant close-in planets, they would also constraint the eccentricity of the orbit and thus the tidal thermal history of the planet.

The value that we determine for the sky-projected angle between the stellar spin and the planetary orbital axis is compatible with  zero ($\beta = 11^{+14}_{-18}$ deg). This good alignment was observed for ten other close-in giant planets (see Winn 2008 and references therein), while a possible misalignment was observed only for the planet XO-3 (H\'ebrard et al. 2008). Together, these results favor migration via tidal interactions with a protoplanetary disk (Lin et al. 1996) as the dominant  mechanism of planetary migration, because it should preserve spin-orbit alignement  (Ward \& Hahn 1994) contrary to migration via planet-planet scattering (Rasio \& Ford 1996) or Kozai cycles (Fabrycky \& Tremaine 2007). 

\begin{acknowledgements} 
We thank the ESO La Silla staff for their support during the CORALIE and HARPS observations. The HARPS consortium is gratefully acknowledged for contributing to some of the observations presented here.
\end{acknowledgements} 

\bibliographystyle{aa}
{}
\end{document}